\documentclass{aa}

\def\lea{\mathrel{<\kern-1.0em\lower0.9ex\hbox{$\sim$}}}
\def\gea{\mathrel{>\kern-1.0em\lower0.9ex\hbox{$\sim$}}}
\usepackage{times}
\usepackage[]{graphics}
\usepackage[]{psfig}

\begin{document}

\thesaurus{03.13.2; 11.06.1; 11.09.1; 11.19.2; 11.19.4; 11.19.5}

\title{Resolved Stellar Populations of Super-Metal-Rich Star Clusters
       in the Bulge of M31}

\author{P. Jablonka\inst{1}
\and F. Courbin\inst{2}
\and G. Meylan\inst{3}
\and A. Sarajedini\inst{4}
\and T.J. Bridges\inst{5}
\and P. Magain \inst{6}}

\offprints{Jablonka}

\institute{DAEC-URA 8631, Observatoire de Paris-Meudon,\\ Place Jules Janssen,
92195 Meudon, France 
\and Universidad Cat\'olica de Chile, Departamento de Astronom{\' \i}a
y Astrof{\' \i}sica,\\ Av. Vicu\~na Mackena 4860, Casilla 306, 
Santiago 22, Chile
\and European Southern Observatory, Karl-Schwarzschild-Strasse 2,
D-85748 Garching-bei-M\"unchen, Germany
\and Astronomy Department, Wesleyan University, Middletown, CT 06459 USA
\and  Anglo-Australian Observatory, P.O. Box 296,
Epping, NSW, 1710, Australia 
\and Institut d'Astrophysique, Universit\'e de Li\`ege, 5 Avenue de
Cointe, B-4000 Li\`ege }

\mail{pascale.jablonka@obspm.fr}
\date{Received / Accepted}
\maketitle

\begin{abstract}

We have applied the MCS image deconvolution algorithm (Magain, Courbin
\& Sohy 1998) to HST/WFPC2 $V$, $I$ data of three M31 bulge globular
clusters (G170, G177, and G198) and control fields near each cluster.
All three clusters are clearly detected, with an increase in stellar
density with decreasing radius from the cluster centers; this is the
first time that stars have been resolved in bulge clusters in the
inner regions of another galaxy.  From the RGB slopes of the clusters
and the difference in $I$ magnitude between the HB and the top of the
RGB, we conclude that these three clusters all have roughly solar
metallicity, in agreement with earlier integrated-light spectroscopic
measurements.  Our data support a picture whereby the M31 bulge
clusters and field stars were born from the same metal-rich gas, early
in the galaxy formation.

\end{abstract}

\section{Introduction}

Stellar  populations are  used   as tools   to probe   star  formation
histories in galaxies.  The bulge of our Galaxy, together with that of
our Local Group galaxy  companion M31, have colors, metallicities, and
kinematics which are typical  of early type  spiral  bulges.  Although
our knowledge about bulges has improved during the last decade, little
is known about their formation and evolution, especially in connection
with the other  components   of galaxies  (Freeman  1993).   There is,
however,  recent evidence that  bulges originate, on rather short time
scales,  during   the very first   phases   of galaxy  formation.  For
example,  in our Galaxy,   the  analysis of Color-Magnitude   Diagrams
(CMDs) of  star clusters located within five  degrees  of the Galactic
center has revealed a metallicity distribution similar  to that of the
surrounding Galactic field stars  (Barbuy et~al.  1998).  In addition,
the metal-rich bulge clusters have been demonstrated  to have the same
age as the  inner-halo   metal-rich old globular  cluster   47~Tucanae
(Ortolani et~al. 1995).

The  globular cluster system of  M31, which is about  twice as rich as
the  Galactic one,  is  among the   most studied  cluster   systems in
external galaxies (Harris  1991).  However, our knowledge comes mainly
from  the  photometric and/or spectroscopic  integrated  properties of
these clusters.  A  few pioneering attempts  have been made  to obtain
cluster CMDs from the  ground, but even with good  seeing at the CFHT,
Heasley et~al.  (1988) and Christian  \& Heasley (1991) were only able
to reach the upper part of the red  giant branch, without reaching the
horizontal branches of G1 and G219, two of the brightest M31 clusters.

With the advent of the Hubble Space Telescope (HST), the situation has
changed.  The CMDs of 10 globular clusters in M31 have been published,
reaching about one magnitude  below the horizontal branch (Fusi  Pecci
et~al.  1996, Ajhar et~al.    1996, Rich et~al.  1996, Holland  et~al.
1997).   These clusters are mainly located  in  the halo of M31, where
contamination  from   foreground stars  and  the  M31  stellar disk is
minimized.  The closest to  the center are G108  (with [Fe/H]=$-$0.94)
and G280 (with [Fe/H]=$-$0.4), located  at 19.2\arcmin\ = 4.81 kpc and
18.4\arcmin\ = 4.59 kpc from the M31 center, respectively.

Recently, Jablonka  et    al.    (1992, 1998)    obtained   integrated
spectrophotometric  observations for a sample  of globular clusters in
the  bulge of  M31,  looking  for   possible  extreme cases  of  metal
enrichment, and  consequently naturally investigated the inner regions
of M31.  The two clusters G170 and G177  belong to their sample.  G177
exhibits metallic absorption  features which are   as strong as  those
characterizing the central  regions  of elliptical galaxies; with  its
higher-than-solar metallicity,  this  cluster  clearly challenged  the
conventional  view   of old    globular  clusters as metal-poor  objects,
although such metal-rich clusters seem to be rare.  Another cluster,
G198, as super-metal rich as G177 and at  about the same distance from
the M31 galactic center,  had been identified  by Huchra et al. (1991)
in an  independent   spectroscopic  analysis. G170,  located  slightly
further away has about a solar metallicity.

As a first  attempt to learn more  about the outstanding properties of
extremely  metal-rich globular clusters,  we applied  for deep imaging
observations   with  the HST, taking    advantage of the  high spatial
resolution and red  sensitivity of the  Planetary Camera  of WFPC2, in
order to build the CMDs of these three M31 bulge clusters, all of them
located within about six arcmin from the center of this galaxy.  These
cluster data, so  far the closest to  the center of M31, allow insight
into the questions of metallicities, age  and link between bulge field
and  cluster stellar  populations.   Given their  central location and
consequently their very high stellar density,  they challenge the best
observational capabilities.   G170, being slightly  further away  from
the center, is surrounded  by a lower density of  field stars than the
other two  clusters, and was  meant to  be a  relative calibrator.   A
fourth   cluster, G1, has been observed   during the same program; its
properties, viz.  CMD, structural parameters, dynamical mass estimate,
and M/L  ratio, will be     addressed in  another paper  (Meylan    et
al. 2000).

The structure of this paper   is  as follows: Section~2 presents   the
observations   and  data reduction;   Section~3  describes  the  image
deconvolution technique   used; Section~4 analyses  and  discusses the
results,  while Section~5  summarizes the highlights of this study.

\section{Observations}

We  obtained HST  WFPC2 images  with  the F555W ($V$)  and F814W ($I$)
filters, during Cycles~5   and~6 (Program IDs =  5907  and 6477).  Our
targets were  field  and cluster stars  around  three  metal-rich star
clusters  G170, G177, and  G198 in the bulge of  M31  (Jablonka et al.
1992 ; Huchra et al.  1991).  The two fields  around the star clusters
G170  and G177  are located South-West   along the major axis  of M31,
respectively  at 6.1\arcmin\ and  3.2\arcmin\ from the galaxy nucleus;
the third field, around the cluster G198,  is located North-East along
the major axis  at 3.7\arcmin\ from  the galaxy nucleus (Huchra et al.
1991  ;  Hodge 1981).  Adopting  1\arcmin\ =  250~pc  (Rich \& Mighell
1995), these angular separations  correspond to projected distances of
about  1.55, 0.80,  and 0.92  kpc,  respectively.  These distances are
summarized in Columns 2 and 3 of Table~1. Columns 4  and 5 of the same
table provide the $V$ magnitude and $B-V$ color  of each cluster, from
Battistini et~al. (1987).  Fig .1 in Jablonka et al. (1999) gives 
the location of our fields.

Figure~1 displays,  on the  left, for  each of the  three clusters, an
area of 64~$\times$~64 pixels from the original PC frames, centered on
the   clusters.  With  the PC  pixel  size  of  0.045\arcsec/pix, this
corresponds to  squares of 2.88\arcsec\   in size.  Although resolved,
all  three clusters appear extremely  compact, with very steep surface
brightness profiles and extremely crowded and bright cores.  Since all
images   are  slightly saturated  in the  core    of the clusters this
prevents the determination of the  cluster structural parameters.   In
any case,  these three  clusters   are extremely concentrated,   being
probably close to core collapse.

Our data represent  one of the  highest resolution  images obtained so
far of the innermost parts  of M31 for  either cluster or field stars.
While  the high spatial    resolution of  the  HST is   necessary when
studying crowded  fields, its  unfortunate under-sampling limits
the effectiveness of  the instrument in  actually resolving blends  of
stars.  The consequences of various  samplings on the study of crowded
stellar fields is  investigated by Renzini  (1998), who shows by clear
physical considerations that the main limiting factor is the number of
stars   per  resolution  element.    For whatever    available spatial
resolution provided  by a given telescope, the  sampling of the images
as determined by the CCD  still  determines the  number of degrees  of
freedom  (pixels),  as  compared with   the number  of  unknowns (star
intensities).

In the center of M31, the number  of unknowns is close  to that of the
available degrees of   freedom,  making it   difficult to  obtain  the
photometry of individual stars.  In   this context, the photometry  is
affected not only by the photon noise, but  also by systematic errors.
Systematics translate  into biased photometry  and may even depend on
the algorithm  used.  For   example, the  way the  sky  subtraction is
performed   influences  significantly  the   results  in images  where
basically no  single pixel  provides  an accurate estimate  of the sky
brightness.  An algorithm  which treats  the sky background   globally
(i.e., the sky  background is an  image) will usually  not produce the
same photometry as an algorithm which computes a local sky value.  The
sky     level  will  usually     be  {\it  systematically}    over- or
under-estimated, hence leading to biased photometry (such an effect,
is of course  best seen  in faint stars,    close to the sky   level).
Another source  of systematic errors comes  from  the way heavy blends
are handled by the photometry software: an  algorithm which does not
"see"  very   blended stars  will tend to    overestimate  the flux of
multiple point  sources  identified   as one  single   source.   While
correcting   for the net  effect of  such systematics  is  extremely
difficult,  one may still be able   to quantify their magnitude.  For
this  purpose, we   performed our  analysis  with  two very  different
algorithms: the  ALLFRAME procedure developed  by P.   Stetson (1994),
and   the "MCS  image   deconvolution algorithm"  developed by Magain,
Courbin \& Sohy (1998).

Both      ALLFRAME   and  MCS     have     been  run on  the
64~$\times$~64-pixel  (2.88\arcsec~$\times$~2.88\arcsec) sub-areas  of
the original PC frames. The small size of these images has been chosen
to minimize  the CPU consumption of the  MCS method.  The clusters are
centered  and fully included in  these subimages.  Reference sub-areas
in the same  PC frames, hereafter  called control fields, with similar
64~$\times$~64-pixel size, have been chosen  at about 10\arcsec\  away
from the clusters, far  enough to  ensure that  no cluster  star would
intervene, though not too far away, so that the conditions of analysis
remain the same.   The directions between  cluster  fields and control
fields are also  chosen to minimize  any effect induced  by the strong
gradient of the bulge stellar  field.  Table~1 gathers the coordinates
of the cluster and control fields in pixel units of the PC frames.  As
ALLFRAME is now widely   known and since our    use of it  is  already
described in Jablonka et al.  (1999), we mainly focus in the following
section on   the description  of our  use   of the  MCS  deconvolution
algorithm.

\begin{figure}[t]
\resizebox{8.9cm}{!}{\includegraphics{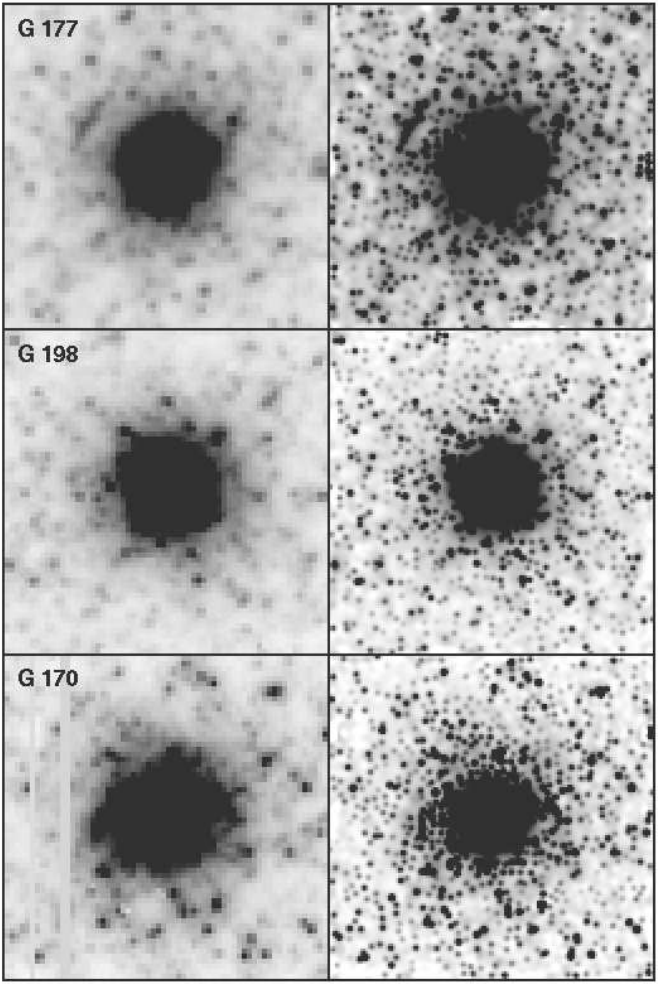}}
\caption[]{\label{fig.diagram}  Left:   These   three panels   display
64~$\times$~64 pixels (2.88\arcsec~$\times$~2.88\arcsec) area from the
original    F814W~=~$I$  PC frames, centered    on  the three M31 star
clusters G177, G198, and G170.  Right: these  three panels display the
corresponding results of the MCS deconvolution algorithm. }
\end{figure}

\section{Image Deconvolution}

The MCS deconvolution algorithm is applied to the combined $V$ and $I$
frames , viz., 4 frames  in each of  the $V$- and $I$-bands.  Figure~1
(left)   displays for  each  cluster   the  64~$\times$~64-pixel  area
centered on  the cluster, from one  of the the $I$-band  images, after
the classical HST pipeline data reduction procedure.  Figure~1 (right)
displays the resulting image after the MCS deconvolution.

The PSF is known to be variable across the WFPC2 field with a scale of
about  ~30\arcsec.  Consequently, the PSF  can be safely assumed to be
constant  over each  of   our  small 64~$\times$~64-pixel areas,   but
different PSFs were computed for  each cluster and each control field.
These PSFs were obtained from the grid of numerical PSF estimates made
available to us by P. Stetson from the Cepheid Key Project.  The F555W
and F814W  instrumental magnitudes were converted  to Johnson-Cousins~
$V$  and $I$ magnitudes, using  the  zero points  in  Holtzman et  al.
(1995)  and the color terms given  by Hughes  et~al.  (1998) for their
Key  Project  WFPC2 data.  No aperture   correction  was applied.  The
resulting color transformation equations are :

\begin{eqnarray}
V & = & V_{MCS}  - 0.045 (V-I) + 0.027 (V-I)^2 + 21. 725 \nonumber\\
  & + & 2.5log (1.987)\nonumber\\
I & = & I_{MCS}  - 0.067 (V-I) + 0.025 (V-I)^2 + 20.839\nonumber\\
  & + & 2.5log (1.987) \nonumber
\end{eqnarray}

Our cluster and control fields fall close to four PSF estimates on the
PSF grid, so that we were able to compute the final  PSFs used for the
deconvolution on a  grid of  pixels a factor  of two  smaller than the
original  pixel size; this is  similar to  the ``drizzling'' procedure
used to over-sample images, using  various dithered frames of the same
field (Fruchter \& Hook 1998).  The final pixel size adopted to sample
the deconvolved images is  then 0.0225\arcsec.  Adopting this  smaller
pixel size is essential  when deconvolving slightly under-sampled data
as  those from WFPC2.    We can then  improve  the  spatial resolution
beyond the  limits  in  principle  imposed by  the  pixel size  in the
original data   frame and  reach  with  2   (smaller) pixels  a  final
resolution equal to 0.045\arcsec~ FWHM, without violating the sampling
theorem.

\begin{table*}
\caption[]{ Cluster Information }
\begin{tabular}{c c c c c c c} 
 	Cluster Name &
        Distance     &
        Distance     &
        $V$       &
        $B-V$     &
        Cluster Field &
        Control Field   \\
	~ &
        [arcmin] &
        [kpc]  &
	[mag]  &
	[mag]  &
        (x1:x2 ; y1:y2) &
        (x1:x2 ; y1:y2) \\
\hline
G177 & 3.2 & 0.80 & 15.91 & 1.31 & 386:449 ; 415:478  & 224:287 ; 510:573 \\
G198 & 3.7 & 0.92 & 15.98 & 1.24 & 322:385 ; 332:395  & 602:665 ; 405:468 \\ 
G170 & 6.1 & 1.55 & 16.45 & 1.23 & 524:587 ; 352:415  & 278:341 ; 320:383 \\
\end{tabular}
\end{table*}

The  deconvolved numerical images are decomposed  into  a sum of point
sources and background, where  the background includes, in the present
application, not   only the sky  background  and all possible extended
sources, but also the diffuse  light from the stellar population still
unresolved in spite of  the  use of the  HST.    The user of the   MCS
deconvolution algorithm  has to  decide  how  many point  sources  are
present in the  deconvolved image: in the  present crowded fields,  we
may miss many of  the faint and/or  overly  blended stars.  While  the
stars missed are obviously not measured as point sources, because they
are not  present in the   point-source part of the  deconvolved image,
they are nevertheless still present in the background component of the
image.    Light pollution of  the stars   actually identified as point
sources is therefore minimized.

Several   consecutive deconvolutions are run in   order to identify as
many stars as possible during  this iterative process. The number
of point    sources to  be  involved  in    the deconvolution   can be
objectively defined by checking the  quality of the residual map (RM),
which  is the difference   between the  original data  frames  and the
deconvolved images  (re-convolved with the  PSF),  in units  of photon
noise (e.g., Courbin et al.  1998).   A good deconvolution, i.e., with
the optimum number of point sources  involved in the fit, should leave
a flat RM with mean value of  1$\sigma$.  Missing one or several stars
results  in significant  residuals,  above the  critical  value of  1.
Adding too  many stars results in over-fitting  of  the data and local
residuals below  1.  We therefore always choose  the minimum number of
stars leading to a  statistically  acceptable RM.  Such  a criterion,
which   works successfully when data   are  correctly sampled, is more
difficult to apply here.   This is true for  all 3 clusters, and it is
clear that we miss  in all cases a  significant fraction of the stars,
especially close to the cluster  centers.  One additional and reliable
criterion we apply to  select the stars to construct   our CMDs is  to
stipulate that  the  detection of  a star  is  genuine only if   it is
present in both the $V$ and $I$ images and if its position is the same
in the two bands, within an error box of 0.25  pixels (in the original
data frames).

Importantly, knowing the PSF  on a grid of pixels  smaller than in the
original  data frames minimizes the  limitation  related to the larger
pixel size of the HST.   The number of  degrees of freedom per unknown
is still fixed by the physical pixel size of  the detector, but we are
able to  resolve closer blends  with  the deconvolution algorithm than
with ALLFRAME.  The typical minimum distance between two point sources
of similar brightness is about 0.3 pixels (in the original data frame)
with the deconvolution, while  ALLFRAME rejects all blends closer than
about 0.7 pixels, equivalent  to 0.37 times the  FWHM of the PSF.  The
systematics  arising  from strong  blends    (see Section  2)   should
therefore  be less severe in  the MCS deconvolution photometry than in
the profile fitting   photometry.  However, the  sky background (which
also included unresolved stars) cannot  be modelled properly using our
data.  In order to test the systematic  errors which may be introduced
by unsecure  sky background  determination, we ran   the deconvolution
several  times,  with   different  initial  parameters  and  smoothing
strengths (see MCS).  We  noted that the  magnitudes measured  for the
faintest stars  can be affected by offsets  of up to 0.3 magnitudes in
both   bands (peak  to    peak  variations), between two   consecutive
deconvolutions using different parameters.  We therefore estimate that
our results  are affected by {\it systematic  errors} (i.e., all stars
systematically  too  bright or  too faint)   of  about 0.07 magnitudes
(1$\sigma$  error, calculated as 0.3  divided by 2 (peak-to-peak total
spread and divided  by 3,  to translate  3  sigma errors into 1  sigma
errors)  in both  bands.  This leads  to  shifts in color   of 0.1 mag
(1$\sigma$). The 1 $\sigma$ error on the positions of the sequences in
our CMDs, due to systematic errors alone are therefore of the order of
0.07 mag in ordinate and 0.1 mag in color.

\section{Results and Discussion}

\subsection{ Star counts }

\begin{figure*}[t]
\resizebox{9.5cm}{!}{\includegraphics{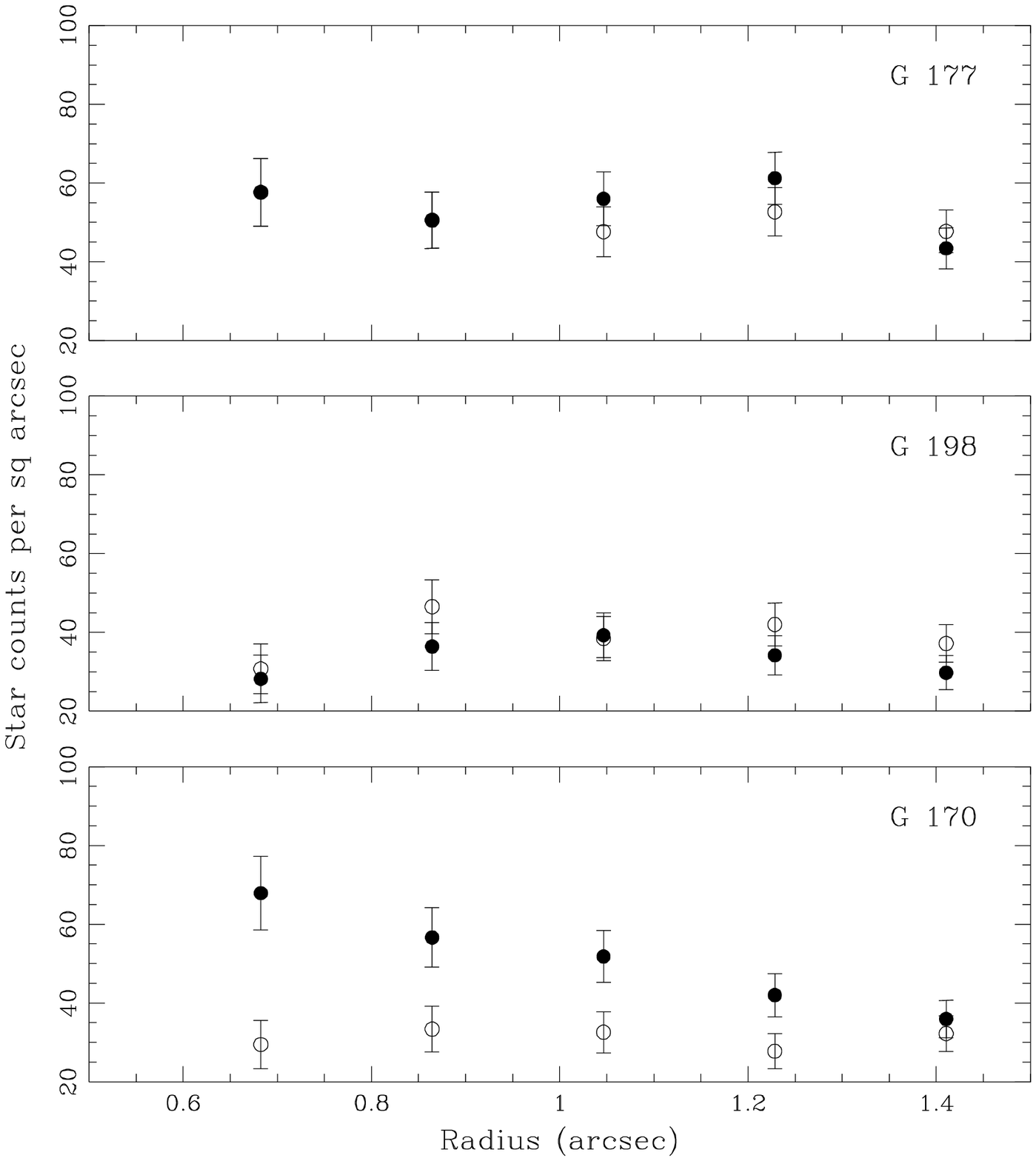}}
\resizebox{9.5cm}{!}{\includegraphics{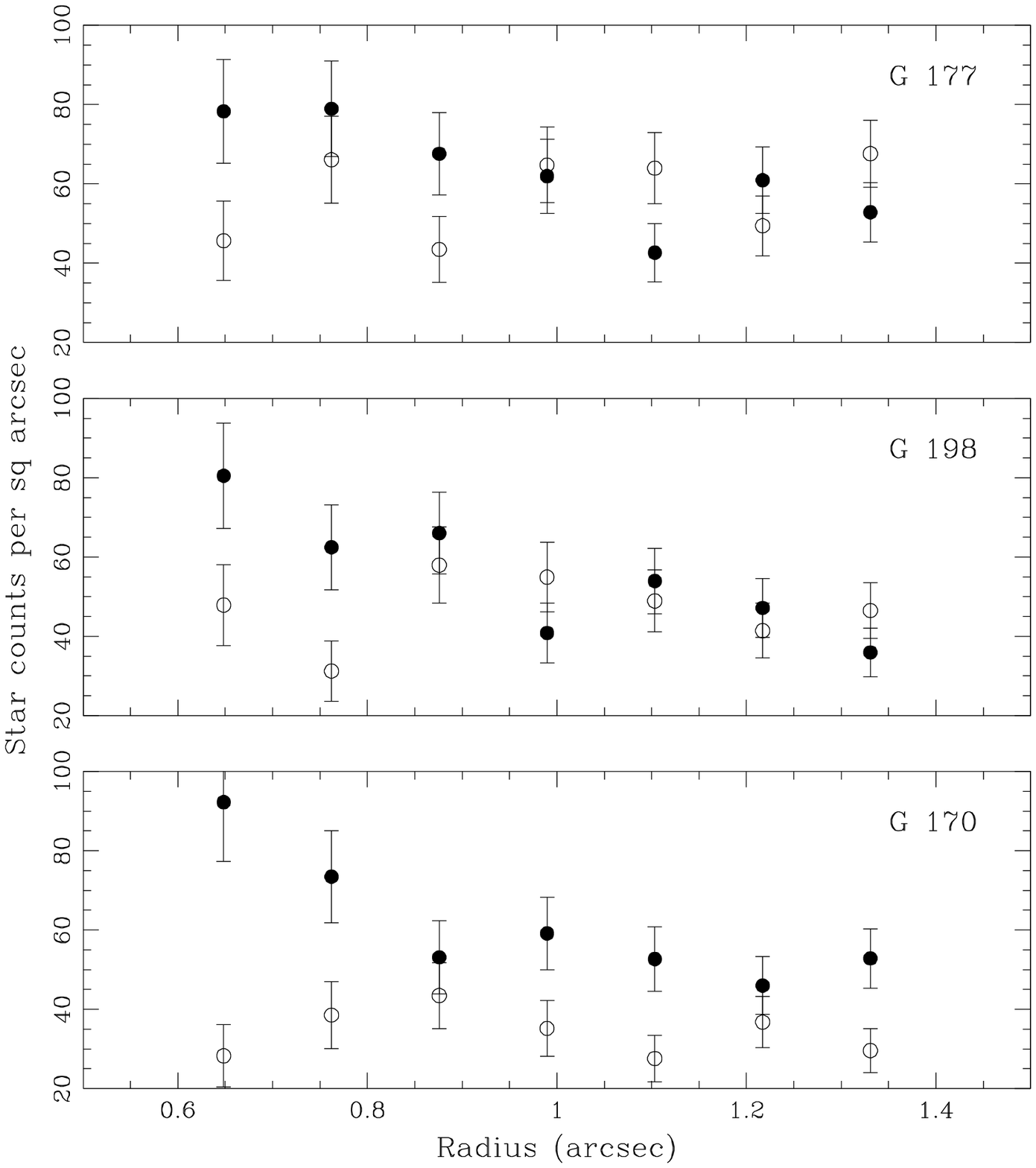}}
\caption[]{\label{fig.diagram}   Radial  star  counts    for all stars
detected with ALLFRAME (left  panel),  and with the MCS  deconvolution
(right  panel),  for  each  of our   three clusters and  their control
fields.  The filled  circles refer to  the  cluster counts, while  the
open circles refer to the control fields.}
\end{figure*}

Figures~2 presents   our      results from ALLFRAME     and  the   MCS
deconvolution.  We plot the number of detected stars per square arcsec
as a function  of the radius, calculated outwards  from the centers of
the clusters or of the reference fields.  For the clusters, we neglect
all   points within  0.55\arcsec\ of the   cluster  centers, where the
crowding  is too high to allow  any reliable results for either method
of analysis.   We also apply  magnitude cuts and  keep only stars with
$V$ $\leq$ 26.5 mag and $I$ $\leq$ 24.5 mag.  These cuts are necessary
if one wants to properly compare the field and the cluster stars which
is one of the main  aims of the present work.   Indeed, the density is
higher in the cluster regions and thus prevents us going as deep as in
the field.  However, only a few stars  are excluded by this criterion.
These  magnitude cuts,  for both  ALLFRAME and  the MCS deconvolution,
correspond to maximum internal errors  (photon noise only) of 0.2 mag.
The  total   errors on  the  photometry  of  the   brighter stars  are
significantly smaller (0.01 mag for the brightest).

In Figures~2,  we   expect to see  the  following  two  effects.   (i)
Detection  of  the clusters: the   star counts for the  cluster fields
should decrease as a function  of increasing radius, as the successive
rings contain fewer and  fewer  cluster stars,  mixed with some  field
stars.  (ii) Detection of  the M31 field  gradient: there should be no
detectable gradient  in star  counts {\it  within} each small  control
field; however the gradient should be observable from one field to the
other,  i.e.,  when considering the  three  different cluster regions,
G177,  G198 and G170  in increasing distance to  the M31 center (Table
1).

\subsubsection{ Detection of the clusters }

The  cluster G170 is very  clearly detected.  The star counts provided
by ALLFRAME  are significantly different  in  the cluster  and control
fields (Figure~2    lower left panel)  and   agree closely  with those
provided by MCS deconvolution  (Figure~2 lower right panel).  There is
a clear increase in  stellar density at  the vicinity of the  cluster,
although with more  stars detected with the  MCS deconvolution.   With
decreasing radius,  ALLFRAME      star counts increase  from     35 to
70~stars/arcsec$^2$, while MCS deconvolution star counts increase from
40 to  95~stars/arcsec$^2$.  At the distance  of G170, the  density of
bulge field  stars is  low enough so  that  everywhere in the  cluster
field  the  density of  detected stars is  higher than  in the control
field.   If  one consider the MCS   counts, the control field contains
about 30  stars  per square  arcsec.   The  outskirts of  G170 have  a
stellar density of $\sim$ 50 stars per square arcsec, which means that
there  are  nearly as  many cluster stars   as field stars.  From 0.75
\arcsec~ inwards, the number of stars from the cluster represent
2/3 of the total stellar population.

In the case of G177  and G198, the  situation is less favorable.  With
ALLFRAME, the star counts of the cluster fields remain compatible with
those of the control fields  (Figure~2 upper and middle left panels).
With  the MCS deconvolution  (Figure~2 upper and middle right panels),
the star counts  increase within 1\arcsec.   There,  the cluster stars
represent about  half  of the  total stellar  population.   This means
that, although quite conspicuous on the original frames, stars in G177
and in G198  were   not detected  by  the  reduction  procedure  which
employed  ALLFRAME  and   are    marginally   detected  by  the    MCS
deconvolution.  The ability of  software  to find stars  depends  very
much on how it can cope with poor sampling and on how well it performs
the    background/point source   separation,  especially  in   crowded
environments.  We  do not detect the presence   of G177 and  G198 with
ALLFRAME, as the confusion limit has  been reached for this technique;
thus  the number of detected stars  may even fall below the background
value, and  stays basically at  the  same level, independently  of the
distance from the cluster center.

\subsubsection{ Detection of the M31 field gradient }

This gradient  can  be  detected  in the control   fields  and  in the
outermost points of the cluster fields when considering the mean level
of star counts.  From both ALLFRAME and MCS deconvolution star counts,
the average density of  field stars in clusters  and control fields is
about 30~stars/arcsec$^2$   around G170,   about   40~stars/arcsec$^2$
around G198,  and 55~stars/arcsec$^2$ around G177.   In both methods of
analysis, the mean density of field stars increases significantly when
considering the fields close to G170, G198, and G177, respectively, as
expected from their locations closer and closer to the galaxy center.

From these considerations, given the success  of the MCS deconvolution
in detecting the cluster stars of G177  and G198, we discuss below
only the photometry resulting from this procedure, in the form of CMDs
for each of the three clusters and control fields.

\subsection {The photometry}

Figure~3 displays  the ($I$ vs.  $V-I$)   CMDs resulting from the  MCS
deconvolution, for both the control (left) and cluster (right) fields.
In all six  panels, we clearly   see the field  and cluster  red giant
branches (RGB).   The CMDs extend to  slightly below  the level of the
horizontal branch  (HB) at $I$   = 24.35, though  we  do not have many 
HB stars due to our general completeness limit  (see Jablonka et al
1999) and  to the   poor statistics in  such   small regions. 

In the CMDs shown in the right panel of Figure~3, the RGB sequences of
the clusters G177 and G198 are  slightly bluer by $\sim$ 0.15 mag than
those of their respective  control fields (left panels).  In addition,
the CMDs between  the clusters also exhibit a  comparable variation in
their mean colors.  This effect is due to the increase in density from
the fields to the clusters or from one mean location to another closer
to the M31 center and  is especially visible between the G170 location
and the  others.  The effect of  crowding leads to blends  of stars as
already discussed by means  of artificial star experiments in Jablonka
et al.  (1999). As illustrated in Figure~4, while the faint and bright
ends  of the  luminosity functions  are not  modified, when  moving to
higher star  densities, the bulk  of stars are shifted  towards higher
luminosities  by blends  of intermediate  luminosity stars.   Since we
observe  intrinsically  red  stellar  populations, confusion  is  more
pronounced  in $V$ than  in $I$,  so that  more unresolved  blends are
faced in $V$,  due to shallower contrast (i.e.,  dynamic range), hence
leading  to bluer  colors.   As  the brightest  end  of the  magnitude
distribution is not affected, the  location of the reddest part of the
RGB is  safe and  allows analysis of  age and abundance.   We estimate
that these  systematic effects may  translate into errors as  large as
0.3 mag in  $V-I$ at $I\sim 24.5$ mag (see section  3), but should not
be larger than 0.1 mag at $I\sim 22-22.5$ mag, at a 3$\sigma$ level.

\begin{figure}[t]
\resizebox{9.0cm}{!}{\includegraphics{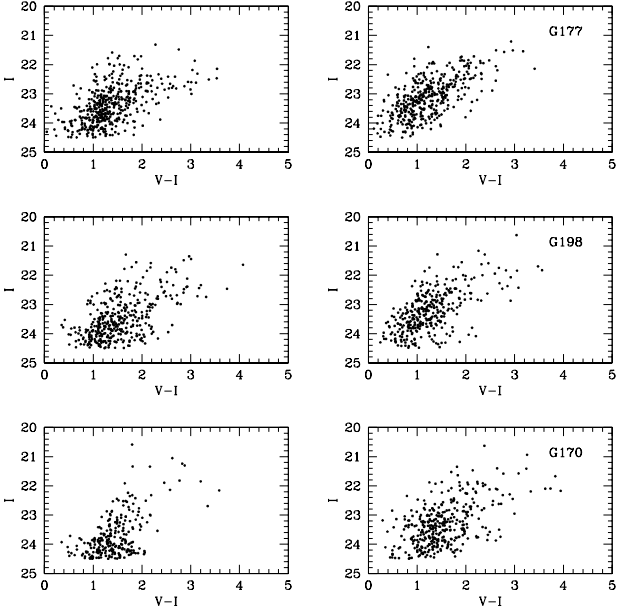}}
\caption[]{\label{fig.diagram} CMDs for all  three control  (left) and
cluster (right) fields.  The 1-$\sigma$ errors in $I$ are of the order
of 0.15 mag between 23.5 and 24.5 mag, 0.02  mag between 22.5 and 23.5
mag, 0.005  mag between 21.5 and  22.5 mag.  The  1-$\sigma$ errors in
$V$ are of the  order of 0.2  mag between 25.5 and  26.5 mag, 0.07 mag
between 24.5 and 25.5 mag, 0.01 mag between 24.5 and 23.5 mag. }
\end{figure}

Following the  approach of Ortolani  et al. (1991),  we derive the RGB
slopes of our clusters in $I$ and $V$ (see Table~2).  We also indicate
the differences in  $I$ magnitude between  the Horizontal  Branch (HB)
level  and  the  top  of the RGB  (the  brightest  stars)  which  is a
complementary indicator of metallicity  (Barbuy  et al.  1997).    For
G170 all stars in the 64x64 pixels frame  are included, while for G177
and G198,  only  stars within  a 1$\arcsec$  radius  are used  in  the
calculation, the  region   where the  presence  of  cluster  stars  is
unambiguous.  The  horizontal branch level  for the  clusters has been
taken to  be equal to that of  the field.  Fusi Pecci   et al.  (1996)
give  M$_V$$^{HB}$  $\alpha$ 0.13 [Fe/H].   In  our  case, even if the
clusters were less metallic than the field, it would be hardly by more
than 0.6~dex, otherwise we would detect brighter red stars than we do;
thus the change in magnitude for the HB  stars going from the field to
the  clusters, or even from  one cluster  to  another would be smaller
than our photometric precision.  The mean magnitude of HB stars at $I$
$\sim$ 24.35 mag has been measured from the luminosity function of the
entire PC frame around G170.  For comparison  and ranking of the three
M31  clusters, we   indicate   the values  derived  for   NGC~6528 and
NGC~6553,  bulge clusters in   our Galaxy, both at [Z/Z$_{\odot}$]=0.0
(Ortolani et al. 1995; Barbuy et al. 1999) and  for the Galactic thick
disk cluster 47 Tuc.  The number of stars on the  red extension of the
giant branch of  G177 ($V-I$  $\ge$  2.5 mag)  in the inner  1\arcsec~
radius is  too low to give  reliable slopes or  magnitude differences.
Indeed, given the small  areas considered, poor statistics prevents us
from   measuring the  RGB slope for   G177.   All three clusters  have
previous spectroscopic determinations of their global metal abundance,
[Z/Z$_{\odot}$] = 0  and 0.3, for  G170 and G177  and [Fe/H] =0.09 for
G198  (Jablonka et  al.   1992,  Huchra  et  al. 1991).   The  present
estimates,  although crude, are  in very good  agreement, as they rank
the clusters at the level  of  NGC~6553 and  NGC~6528.  This gives  us
enough  confidence in our photometry  to pursue the comparison between
the cluster and field stellar population.

From the various CMDs in Figure~3, and given our error bars and
systematic biases discussed above due to image crowding, we are led to
conclude that there is no significant difference between the cluster
and the mean field stellar populations; the clusters are indeed in the
M31 bulge and are not seen in projection.  Our data indicate a
formation of the M31 bulge clusters from the same material as that of
the field stars, at an early epoch in the formation of M31. 

\begin{table*}
\caption[]{RGB slopes and magnitudes}
\begin{tabular}{l c c c c} 
	Cluster  &
	[$Z/Z_{\odot}$]  &
         $\Delta I $ /  $\Delta (V-I)$  &
         $\Delta V$ /  $\Delta (V-I) $  &  
         $\Delta I_{HB}^{RGB}$    \\
\hline
47 Tuc    & $-$0.4 & $-0.24$ $\pm$ 0.07  & 0.77 $\pm$ 0.07  & 3.46 \\
NGC 6528 & 0.0    & $-$0.06 $\pm$ 0.07  & 0.82 $\pm$ 0.07  & 2.8  \\
NGC 6553 & 0.0    & $-$0.02 $\pm$ 0.10  & 0.92 $\pm$ 0.2   & 2.7  \\
G170     &$\sim $0.0 & $+$0.01 $\pm$ 0.08  & 1.00 $\pm$ 0.08  & 3.   \\
G177     &$\sim $0.0 &       --            &       --         & 2.8  \\
G198     &$\sim $0.0 & $+$0.01 $\pm$ 0.07  & 1.00 $\pm$ 0.10  & 3.1  \\ 
\hline
\end{tabular}
\end{table*}

\begin{figure}[t]
\resizebox{9.0cm}{!}{\includegraphics{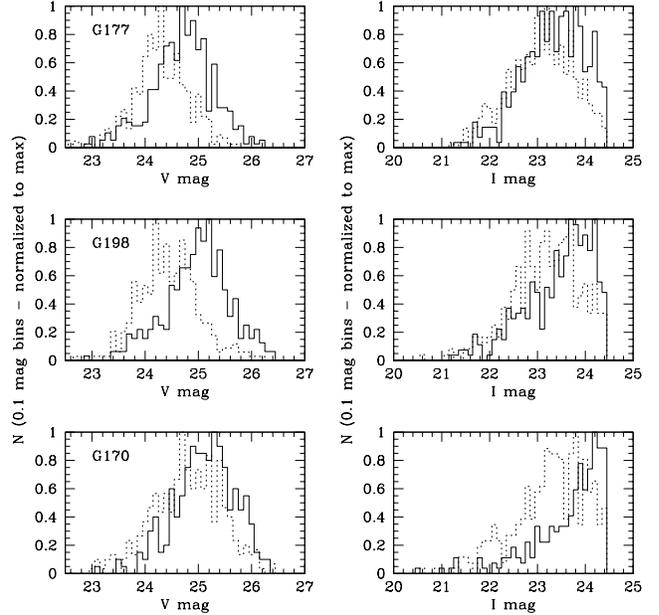}}
\caption[]{\label{fig.diagram} $V$ (left)  and  $I$ (right) luminosity
functions for control and cluster  fields.  Solid lines give the field
luminosity    functions and the  dotted   lines the cluster luminosity
functions.}
\end{figure}

\section{Conclusions}

We  have applied the  MCS  deconvolution algorithm  to three M31 bulge
globular clusters.   This  deconvolution appears as   a very efficient
method in such extremely dense  regions, but still rather demanding in
terms   of computing  time, so   we  have been limited  to small
regions.   However, our initial results   are already very  promising.
Our radial star counts for the cluster and control fields clearly show
that we  have resolved  the  clusters:  this is  the first  time  that
cluster  stars have been  resolved in  the bulge of  another galaxy so
close to the center.

From the deconvolved photometry, we  clearly see the cluster red giant
branches, though  we do not detect their  horizontal branches.  Taking
into account  systematic biases introduced  by crowding, there  are no
significant  differences  between the  CMDs  for  the three  clusters.
Moreover,  there is  no apparent  difference between  the  cluster and
their  respective control  field CMDs.   From the  RGB slopes  and the
difference in $I$  magnitude between the HB and  RGB, all three clusters
have roughly  solar metallicity, making  them similar to  NGC~6528 or
NGC~6553 in  our Galaxy; by  inference, the same  is true of  the mean
field population.

We thus conclude that the M31 bulge clusters and field stars originate
from the  same material.  They are   representative of old, metal-rich
populations.  Similar results have been  found in our Galaxy, where it
becomes more and more  apparent that the metallicity  distributions of
field  stars  and globular clusters  in  the bulge are identical
(Ortolani et al. 1995; Barbuy et al. 1998).  From  a detailed study of
the element ratios for  two stars in  NGC  6553, Barbuy et al.  (1999)
conclude that  the Galactic bulge underwent  rapid star  formation and
chemical  enrichment.   Certainly  our  work  corroborates  this view.
Bulges appear to be old, metal-rich  systems, similar in many respects
to elliptical galaxies.  There is mounting evidence that most stars in
ellipticals and bulges formed at high redshift z$_f$ $>$ 3 (see review
by Renzini \& Cimmati 1999).

Unfortunately, we have pushed HST  to its limits, and  M31 seems to be
the furthest  galaxy  for which we  can use  HST in  this way.  Higher
spatial   resolution  studies of  bulge populations   in  M31 and more
distant   galaxies   awaits adaptive   optics  on 8--10m  ground-based
telescopes, and eventually the NGST.

\begin{acknowledgements}

We would like to thank Peter Stetson for kindly sharing with us the 
Cepheid Key Project PSFs, and for his support with ALLFRAME.
Ata  Sarajedini was supported   by the National Aeronautics and  Space
Administration  (NASA) grants  HF-01077.01-94A, GO-05907.01-94A,   and
GO-06477.02-95A  from the Space Telescope  Science Institute, which is
operated by the Association of Universities for Research in Astronomy,
Inc.,   under  NASA     contract    NAS5-26555. Fr\'ed\'eric   Courbin
acknowledges financial support  through Chilean grant FONDECYT/3990024
and additional funding from the European Southern Observatory.
\end{acknowledgements}

\end{document}